\documentclass[aps,prl,twocolumn,showpacs,10pt,preprintnumbers,nofootinbib]{revtex4-1}

\usepackage[margin=7pt,justification=centerlast]{caption}

\usepackage{amsmath}
\usepackage{amsfonts}
\usepackage{amssymb}
\usepackage{graphicx, rotating}
\usepackage{epstopdf}
\usepackage{epsfig}
\usepackage{latexsym}
\usepackage{mathtools}
\usepackage{MnSymbol}
\usepackage{graphicx}
\usepackage{color}
\usepackage[dvipsnames]{xcolor}
\usepackage{amsmath,amssymb}
\usepackage{multirow}
\usepackage{hhline}
\usepackage{array,makecell}
\usepackage[utf8]{inputenc}
\usepackage{arydshln}
\usepackage{ mathrsfs }
\usepackage{orcidlink}

\usepackage{tikz-feynman}
\tikzfeynmanset{warn luatex=false}

\usepackage{slashed}
\usepackage{hyperref}
\hypersetup{colorlinks, citecolor=black, linkcolor=black, urlcolor=bluscuro}
\definecolor{rossos}{cmyk}{0,1,1,0.55}
\definecolor{bluscuro}{rgb}{0.15, 0.2, .85}
\definecolor{bluchiaro}{cmyk}{1,.3,0.,0.1}
\definecolor{verdescuro}{rgb}{0.3,0.8,0.3}

\newcommand{\be}{\begin{equation}}
\newcommand{\ee}{\end{equation}}          
\newcommand{\bea}{\begin{eqnarray}}
\newcommand{\eea}{\end{eqnarray}}
\newcommand{\bc}{\begin{center}}
	\newcommand{\ec}{\end{center}}

\newcommand{\TeV}{\,\mathrm{TeV}}
\newcommand{\GeV}{\,\mathrm{GeV}}

\def\hp{{h^\prime}}
\def\cale{{\mathcal{E}}}

\def\MG{\texttt{MadGraph}}
\def\PT{\textsc{Pythia} 8}

\begin{document}

\title{Energy Correlators of Hadronically Decaying Electroweak Bosons}

\author{Lorenzo Ricci\orcidlink{0000-0001-8704-3545}}
\affiliation{Theoretical Particle Physics Laboratory (LPTP), Institute of Physics, EPFL, Lausanne, Switzerland}
\author{Marc Riembau\orcidlink{0000-0002-9842-2425}}
\affiliation{Theoretical Particle Physics Laboratory (LPTP), Institute of Physics, EPFL, Lausanne, Switzerland}

\begin{abstract}
\noindent 
{Energy correlators are field-theoretically clean and phenomenologically valuable probes of QCD dynamics. 
We explore the possibility of using the information encoded in the energy correlators of a hadronically decaying electroweak vector boson in order to extract its full decay density matrix. 
The kinematics of the one- and two-point energy correlators can indeed discriminate between longitudinal and transverse modes and reveal the interference pattern between different vector polarizations.
Such observables improve the sensitivity to microscopic new physics affecting the production rate of the different helicities. We assess the impact on higher dimensional EFT operators in simple scenarios. 
}

\end{abstract}

\maketitle

\medskip

\textit{\textbf{Introduction}}

Scattering at high energies reveals the true nature of massive Standard Model (SM) particles: they are a mixture of fields with different dynamics.
This difference is commonly exacerbated in presence of some putative microscopic physics, out of which the SM emerges. 
For instance, such dynamics might have a role in the EW symmetry breaking mechanism and couple predominantly to longitudinal vector bosons. 
It is therefore of utmost importance to develop a program centered on observables capable to discriminate the various helicity configurations.

Observables of this kind were proposed long ago for $WW$-production at LEP \cite{Duncan:1985ij,Hagiwara:1986vm,Hagiwara:1989mx}, or for $t\bar{t}$-production at Tevatron \cite{Czarnecki:1994pu,Mahlon:1995zn,Mahlon:1997uc}, 
and are a substantial part of the BSM precision program at the LHC \cite{Pretz:2018bze,Goncalves:2018ptp,Banerjee:2019pks,Azatov:2019xxn,Banerjee:2019twi,Chen:2020mev,Banerjee:2020vtm}.
Advantages of using such observables to constrain BSM physics are particularly clear in the context of Effective Field Theories (EFT), where microscopic physics is encoded in higher dimensional operators. At high energy, a given operator modifies only certain helicity amplitude, which is often suppressed in the SM  \cite{Azatov:2016sqh,Franceschini:2017xkh}.
As a consequence, the interference between the SM and EFT amplitudes gets suppressed, which translates to a poor sensitivity and that might compromise their interpretation within a consistent EFT framework \cite{Farina:2016rws,Contino:2016jqw}.
Recovering the sensitivity requires the detailed analysis of the differential kinematic distributions of the final states. For instance, the strategy proposed in \cite{Panico:2017frx} to measure the interference between BSM and SM was later used in \cite{CMS:2021cxr}, improving the linear constraints by an order of magnitude.

In the particular case of diboson production, while many studies exist for leptonic decay products of vector bosons, the case of hadronic decays is less explored and substantially more challenging. Existing results rely on jet substructure and, recently, on machine learning in order to discriminate among longitudinal and transversely polarized vector bosons \cite{Almeida:2008yp,Thaler:2010tr,Larkoski:2014gra,Larkoski:2015kga,De:2020iwq,Kim:2021gtv}.

In this \textit{Letter},  instead, we explore the possibility of using energy correlators (EC) in hadronically decaying $W$ and $Z$ bosons in order to extract information about their helicity. This means not only a better discrimination of signal versus background in searches for specific EFT deformations, but also to design observables sensitive to interferences among different helicity states.  We follow this idea through two simple paths. First, we show how one-point EC can disentangle purely longitudinal and transverse decays. Then we prove that two-point EC can reveal the interference pattern among different helicities. In particular this enhances the sensitivity to EFT deformation whenever SM and BSM amplitudes are produced in different helicity configurations.  We remark that most of our analysis will be theoretical and illustrative. We will explore relevant scenarios with a simplified analysis to grasp the advantages of using polarization-sensitive observables and comment on how to export our results to the LHC framework.

\medskip
\textit{\textbf{Energy correlators \& decay density matrix}}

In general, the amplitude for the production and decay of a vector boson is described, under the narrow-width approximation, by the product of the production and decay amplitudes $\mathcal{A}^{hard,V}_{h}$ and $\mathcal{A}_{h,X}$ that describe the production and decay of a vector boson $V$ with helicity $h$. The cross section involves then a double sum over the helicities of the vector in the amplitude and its conjugate. Therefore, it is convenient to write the cross section as the trace of the product of two matrices
\be
d\sigma = d\rho^{hard,V}_{\hp h} d\rho^V_{\hp h},
\ee
where $d\rho^{hard,V}_{\hp h}$ is the so-called density matrix for the ``hard'' production of the vector $V=W^\pm,Z$  and $d\rho^V_{\hp h}$ is the --fully differential-- density matrix for its decay. The indexes $h,\hp=+,-,0$ run over the vector boson helicities.
Common observables targeting to measure new physics in $d \rho^{hard,V}$, like bump searches or many ``high-$p_T$'' EFT probes, ignore the full information carried by the decay products, integrating over their kinematic variables. 
For instance, since $d\rho^V_{\hp h}\propto e^{i(h-\hp)\phi}$ the decay matrix becomes diagonal if the observable is inclusive in the azimuthal angle $\phi$ and one becomes sensitive only to the diagonal entries of $d\rho^{hard,V}_{\hp h}$, i.e. one loses the information on the interference between different polarizations. This information can be in principle recovered from the kinematic distribution of the decay density matrix  $d\rho^V_{\hp h}$. 

At leading order in the weak coupling constant, the vector boson decays into a $q\bar{q}$ pair, which then interact, radiate and evolve according to QCD dynamics, resulting in a jet of hadrons.  It is our goal to find sensible observables that can probe the decay density matrix $d\rho^V_{\hp h}$ into hadrons in order to dissect the nature of the microscopic physics mediating the hard process. An approach to define such observables consists in studying jet substructure in terms of correlation functions of light-ray operators~\cite{Sveshnikov:1995vi,Ore:1979ry,Chen:2020vvp}
\begin{align}
\cale_i =\text{lim}_{r \rightarrow \infty} \int_0^{+\infty} dt\, r^2 n^i T_{0i} (t,r \vec{n})\,,
\end{align}
where $T_{\mu\nu}$ is the stress-energy tensor.
In particular, we focus on the so-called light-ray density matrix \cite{Lee:2022ige}, defined as the vacuum expectation value
\begin{align}
\begin{aligned}
d\rho^V_{h \hp}[\left\{\cale_i\right\}] = \mathcal{N} \int d^4 x e^{i q \cdot x}  \langle \mathcal{O}_h(x) \cale_1\ldots\cale_N \mathcal{O}^{\dagger}_{h'}(0)\rangle
\end{aligned}\,,
\label{Eq:NpointCorrelator}
\end{align}
where $\mathcal{O}_h$ is some operator exciting the QCD vacuum and $\mathcal{N}$ is just a normalization constant. Physically, $d\rho^V_{\hp h} [\cale_i\ldots\cale_N]$ measures the energy deposited in $N$ calorimeters placed at infinity in some directions $\vec{n}_i$ in presence of a source. In our case, the source $\mathcal{O}_h$  is a \textit{boosted} on-shell vector of helicity $h$ and four-momentum $q$ in the lab. frame, ``injecting'' a $q\bar{q}$ current in the correlator of eq.~\eqref{Eq:NpointCorrelator}
\begin{align}
\mathcal{O}_h(x) = \left( \bar{q}\gamma_{\mu} (g_LP_L+g_RP_R) q \right) (x) \, \varepsilon^{\mu}_h \,,
\end{align} 
Being Infrared and Collinear Safe, energy correlators are theoretically clean and can be safely computed with the usual rules of perturbative QCD. 
Moreover, they possess a series of phenomenologically interesting features. They can be computed and measured on tracks, improving the angular resolution, and the insensitivity to soft radiation mitigates the need of grooming \cite{Chen:2020vvp,Li:2021zcf}.
In the recent years, there has been renewed interest due to new theoretical insights,  allowing a simpler characterization of jet dynamics \cite{Hofman:2008ar,Belitsky:2013xxa,Holguin:2022epo,Komiske:2022enw}.

The $d\rho^V_{\hp h}[\left\{\cale_i\right\}]$ is in general independent of the microscopic dynamics.  However, in the following, we will discuss how the detailed study of energy correlators allows to characterize the diagonal and off-diagonal entries of the decay density matrix, improving sensitivity to new physics encoded in the hard process.

\medskip
\textit{Diagonal entries: Longitudinal and transverse}.---

\begin{figure}
	\centering
	\includegraphics[width=0.8\linewidth]{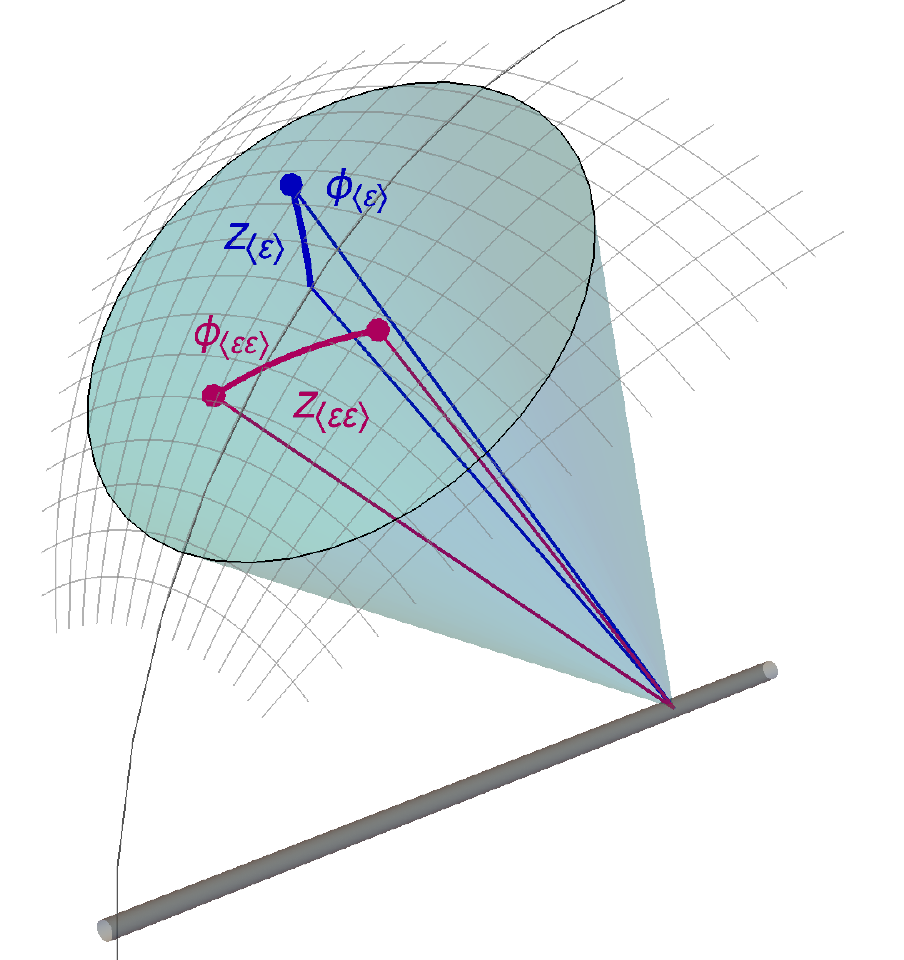}
	\caption{One and two point energy correlator of a hadronically decaying electroweak boson. Dots denote energy measurements from a calorimeter. The one point measures the energy deposit in a calorimeter at a distance $z_{\langle \cale\rangle}$ from the jet axis and azimuthal angle $\phi_{\langle \cale\rangle}$ with respect the plane formed with the beam. The two point correlator depends on similar quantities, but we integrate the distance and orientation of the calorimeters within the jet.}
	\label{fig:ecdiagram}
\end{figure}

The density matrix for the one point energy correlator, that we denote shortly by $\langle \cale \rangle$, is given by the correlator of eq.~\eqref{Eq:NpointCorrelator} with only one insertion of the light-ray operator $\cale_{\vec{n}}$,  pointing toward the direction $\vec{n}$.  For clarity it useful to relate the energy correlator to the more ``phenomenologically friendly'' objects of the weighted cross-section (see for instance \cite{Bauer:2008dt}).  In particular we have
\begin{align}
\langle \cale_{\vec{n}} \rangle =\frac{1}{2 m_V \Gamma_V} \sumint_X \mathcal{A}_{h,X} \mathcal{A}_{h',X}^* \sum_{i\in X}\frac{E_i}{E} \delta \left( \vec{n}_i -\vec{n}\right)\,,
\label{Eq:EC}
\end{align}
where $X$ is a generic multiparticle state to which the vector $V$, of energy $E$, decays. The factor $\frac{E_i}{E}$ weights the energies of the partons in the decay.
Notice that eq.~\eqref{Eq:EC}, or equivalently eq.~\eqref{Eq:NpointCorrelator} with one $\cale$ insertion, depends at most by the two variables parametrizing the direction of $\vec{n}$.  The natural choice are the two angles $0\le \theta < \pi$ and $0 \le \phi< 2 \pi$,  where $\theta$ is the angle respect to jet momentum $q$. The angle $\phi$ is the angle between the line that connects the center of the jet and the calorimeter, and the plane that goes through the jet axis and the beam, see Fig.~\ref{fig:ecdiagram}.
Diagonal entries have a trivial $\phi$ dependence and so they are functions only of $\theta$.

At leading order, the density matrix is determined by the different decay amplitudes $\mathcal{A}_h$ ($\equiv \mathcal{A}_{h,q\bar{q}} $).  In the vector rest frame the latter are  simply proportional to the Wigner-d matrices,
\be
\mathcal{A}_\pm = g_Vm_Ve^{\pm i\phi}\frac{1\pm\cos\theta_*}{2}\,,
\quad
\mathcal{A}_0 = -g_Vm_V \frac{\sin\theta_*}{\sqrt{2}},
\ee
where $\theta_*$ is the angle between the direction of flight of the vector and the helicity-plus fermion.
In the case of the $W$, the coupling is left-handed and $g_W=g$. The $Z$ boson, instead, couples to both left and right currents, with strength $g_Z=\sqrt{2}(gc_\text{W}T^3_f-g^\prime s_\text{W} Y_f)$. 

In the lab frame, the vector boson is boosted up to an energy $E$, which sets a characteristic angular scale $z_\star\equiv m_V^2/E^2$ that controls the kinematics. In this frame, we define $z$ to be the separation between a parton and the direction of the vector boson, $z=\frac{1-\cos\theta}{2}\simeq \frac{1}{4}\theta^2$, see Fig~\ref{fig:ecdiagram}. At leading order in $z_\star$, the energy of a quark is in one-to-one with its separation $z_q$, 
\be
E_q\,=\, E\,\left(1+\frac{4z_q}{z_\star}\right)^{-1}\,\equiv\, E \,x(z),
\ee
where it is convenient to define the quark energy fraction $x(z_q)\equiv x$.
The rest frame angle is related to $z_q$ by  $\cos\theta_*=\frac{1-4z_q/z_\star}{1+4z_q/z_\star}$. The position of the two quarks obeys $z_qz_{\bar{q}}=(\frac{z_\star}{4})^2$, since $z_\star/4$ corresponds to $\theta_*=\frac{\pi}{2}$, so at parton level each event has two energy depositions at each side of $z_\star/4$.
For $z\ll z_\star/4$, the energy of the vector boson is entirely deposited in a single quark since the other quark's energy is redshifted to zero. 
For $z\sim z_\star/4$, the energy is shared among quarks.
In terms of the energy fraction of the positive-helicity quark, the amplitudes in the lab frame at leading order in $z_\star$ are given by
\bea
\begin{aligned}
\mathcal{A}_+ &\simeq g_Vm_Ve^{i\phi}x\,,\quad
\mathcal{A}_- \simeq g_Vm_Ve^{-i\phi}(1-x)\,,\\
&\mathcal{A}_0 \simeq -g_Vm_V \sqrt{2}\sqrt{x(1-x)}.
\end{aligned}
\label{eq:Ampsenfractions}
\eea
Notice that the different helicities have a different behavior, given by the degree of overlap with the $\cos\theta_*\sim 1$ region of the decay amplitude. When measuring the energy correlator, though, we average the position of the calorimeter. Placing it on the minus-helicity fermion is equivalent to $x\to 1-x$ and $\phi\to \pi+\phi$. The quark helicity is not directly measurable, so the energy correlator has the redundancy $\mathcal{A}_h\to (-1)^{h}\mathcal{A}_{-h}$.
Notice also that at leading order in $z_\star$ the kinematics only depend on the quantity $z/z_\star$. This will be of crucial importance, later on, for our discussion on the LHC.

\begin{figure}
	\centering
	\includegraphics[width=1\linewidth]{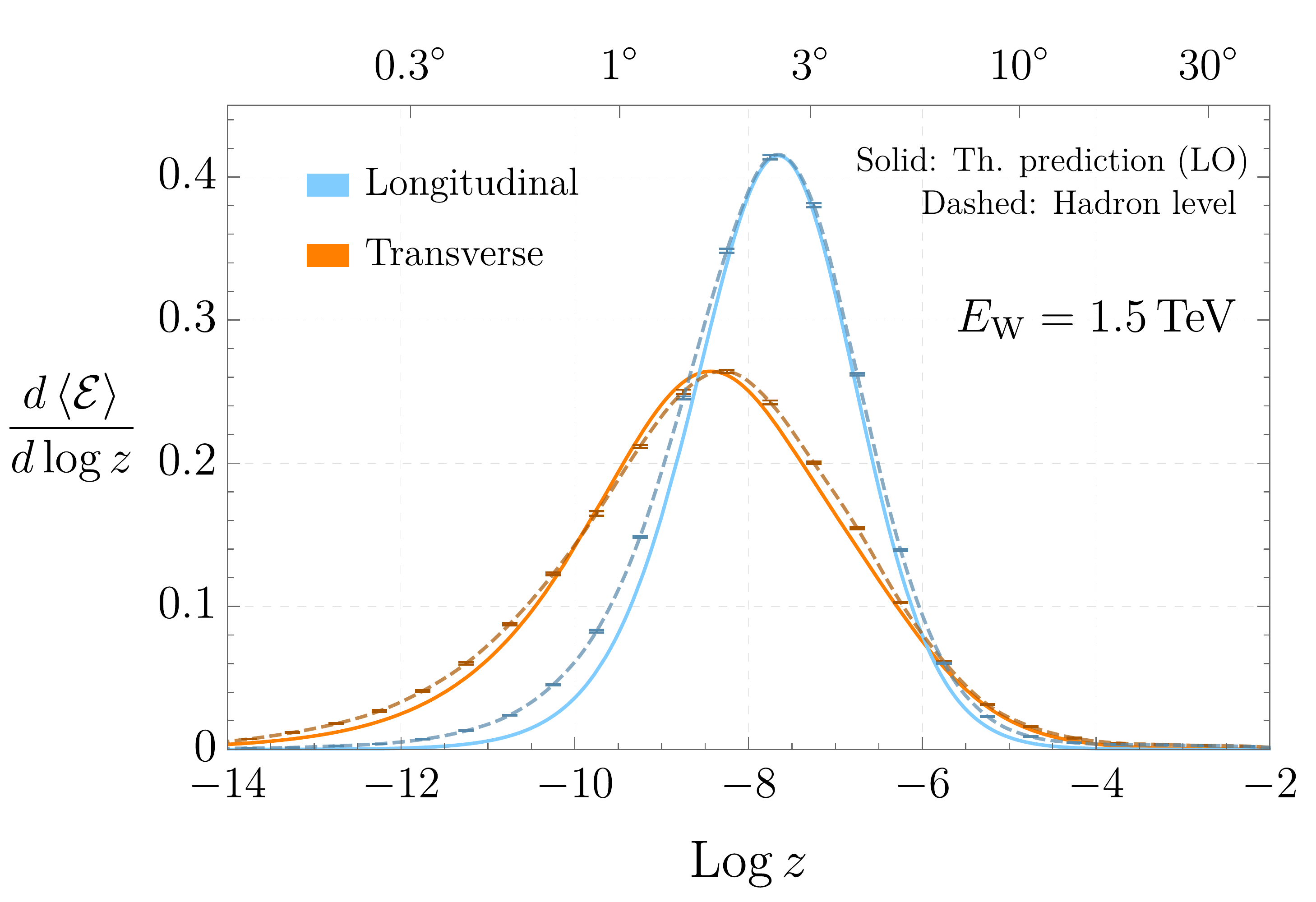}
	\caption{One-point energy correlator of a $W$ jet. The coordinate $z$ is the angular distance between the calorimeter and the center of the jet.}
	\label{fig:EC}
\end{figure}

The transverse and longitudinal terms in the diagonal entries of the density matrix behave differently, since $d\rho^V_{00}\sim x(1-x)$ while $d\rho^V_{TT}\equiv d\rho^V_{++}+d\rho^V_{--}\sim x^2+(1-x)^2$. 
In Fig.~\ref{fig:EC} we show the spectrum of the one-point energy correlator of $d\rho^V_{00}$ (in cyan) and $d\rho^V_{TT}$ (in orange) as a function of $z$ with $\phi$ integrated over, where we fix the $W$ boson energy at $1.5\TeV$. The horizontal lines show the data points with the statistical Monte Carlo error, clearly negligible and in dashed we show the interpolation. The longitudinal spectrum peaks at $z=\frac{1}{6}z_\star$. 
The transverse spectrum is actually the sum of two contributions. The one that dominates by far has the calorimeter placed at the position of the fermion with the same-sign helicity as the $W$ boson helicity, which has a peak at $z=\frac{1}{16}z_\star$. The subdominant contribution has it placed in the opposite-sign helicity quark, which has a much smaller peak at larger $z=\frac{3}{8}z_\star$ and pushes the total transverse peak of Fig.~\ref{fig:EC} to $z\simeq\frac{11}{144}z_\star$. 

In Fig.~\ref{fig:EC} we compare the analytic LO result with the numerical simulation for a monoenergetic ($E=1.5$ TeV) hadronic $W$ jet. Concretely, we simulated through {\MG} \cite{Alwall:2014hca} the semileptonic $W^+ W^-$ production from lepton-lepton scattering. The simulation is further processed via {\PT}  \cite{Sjostrand:2006za,Bierlich:2022pfr} to include the effects of showering and hadronization. The minimum $p_T$ for the $W$ is fixed at $300$ GeV.  
Each event gives a distribution for the correlator, obtained by measuring the energies of the hadrons hitting the detector. In the Figure we show the sum of such distributions, corresponding to the average energy measured by a calorimeter placed at $z$. The LO predictions are normalized to 1, while the {\PT} ones are fixed to have the same-height peak.

The LO calculation gives already a very good prediction for the distributions from \PT, which can be further improved using the higher order corrections.\footnote{Actually, the transverse distribution can be directly imported from \cite{PhysRevD.17.2298,PhysRevLett.41.1585},
by considering their result in a center of mass $\sqrt{\hat{s}}=m_W$, and boosting it by $e^\eta\simeq 2E/m_W$.}

\begin{figure}
	\centering
	\includegraphics[width=1\linewidth]{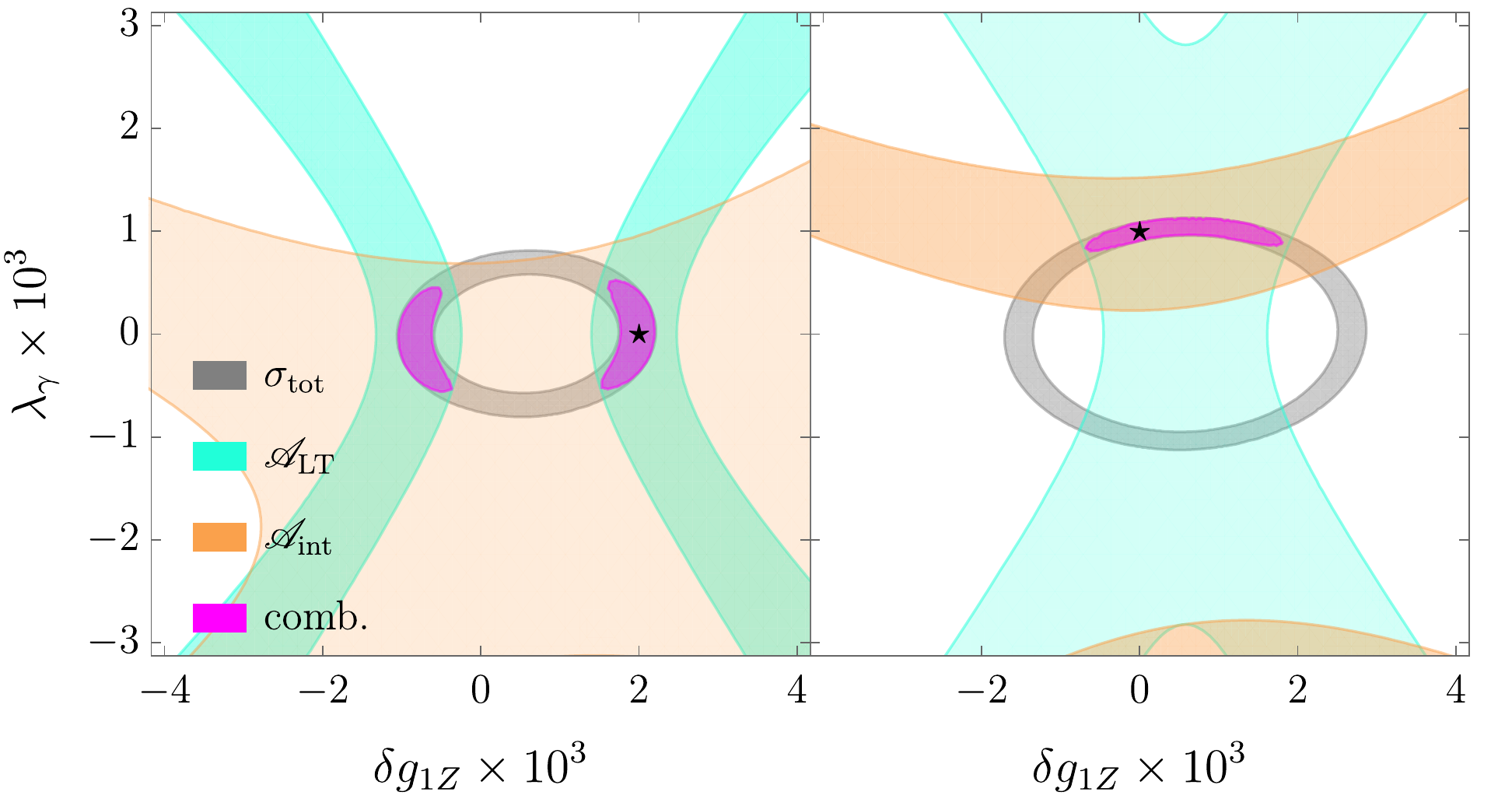}
	\caption{95\%CL contours of different observables of $WW$ production. In the left (right), we assume anomalous production of longitudinal (transverse) $W$-bosons due to non-vanishing $\delta g_{1Z}\,(\lambda_{\gamma})$, indicated by the star.}
	\label{fig:2dconstraints}
\end{figure}

In order to have an estimate of how the differences in the longitudinal and transverse distributions allow to discriminate them,
we define the simple but intuitive \textit{asymmetry} $\mathscr{A}_{LT} =(\mathscr{A}_+-\mathscr{A}_-)/(\mathscr{A}_++\mathscr{A}_-)$, where $\mathscr{A}_\pm$ is the energy accumulated with $z$ greater/smaller than the angle $z^{\text{peak}}$ at which the transverse distribution has a maximum. Approximately, $\mathscr{A}_{LT}\simeq 0$ for a purely transverse sample, while $\mathscr{A}_{LT}\sim 0.5$ for a purely longitudinal one.
At parton level, in the vector boson rest frame, $z^{\text{peak}}$ corresponds to an angle $\theta_*\simeq 58^\circ$. 
In a typical observable where one compares the number of events at each side of a cut, the maximal sensitivity of an excess of longitudinal events over a purely transverse sample corresponds to a cut at $\theta_*\simeq 56^\circ$.\footnote{We remark that the one-point energy correlator measures the average energy density deposited at $z$ in a sample of jet events, and has not to be thought as a probability distribution. However, since $z^{cut}\ll z_\star/4$, at the left of the cut most of the energy is deposited in a single quark and therefore the asymmetry is roughly an event-level counting.}

Let us investigate the impact of the asymmetry in a simple but relevant scenario. In the left of Fig.~\ref{fig:2dconstraints} we consider semileptonic $W^+W^-$ production at $\sqrt{s}=3\,\text{TeV}$ with $p_T(W)>1.2$ TeV and $5/\text{ab}$ of integrated luminosity.\footnote{The choice for the $p_T$ cut is to reduce the large forward SM background improving the EFT sensitivity.} The production of longitudinal modes is considered to be modified by an anomalous triple gauge coupling $\mathcal{L}\supset igc_{\theta_W} \delta g_{1Z}(W^+_{\mu\nu}W^-_\mu-W^-_{\mu\nu}W^+_\mu)Z_\nu$ with  $\delta g_{1Z}=2\times 10^{-3}$ \cite{LHCHiggsCrossSectionWorkingGroup:2016ypw,Degrande:2020evl}.
We interpret the cross section measurement in the 2$d$ parameter space defined by $\delta g_{1Z}$ and $\lambda_\gamma$, the coefficient of $\mathcal{L}\supset i\frac{e}{m_W^2}\lambda_\gamma W^+_{\mu\nu}W^-_{\nu\rho}A_{\rho\mu}$, which instead controls the production of the transverse modes.
The cross section measurement gives the gray contour, and cannot determine the origin of the excess.
The measurement of the asymmetry $\mathscr{A}_{LT}$, in cyan, confirms that the excess comes from an anomalous production of longitudinal modes.  We will comment the rest of the plot later, when discussing the interference terms.

\medskip
We now turn to the two point correlator $\langle \cale_1 \cale_2 \rangle $, defined in a similar way inserting two light-ray operators into eq.~\eqref{Eq:NpointCorrelator}. Potentially, $\langle \cale_1 \cale_2 \rangle $ has a non trivial dependence on four angles, parametrizing the position of $\vec{n}_1$ and $\vec{n}_2$.  In the following, we only consider two angles, being the angular distance $z=\frac{1}{2}(1-\vec{n}_1\cdot \vec{n}_2)$ between the two calorimeters and the azimuthal angle $\phi$ between their line of separation and the scattering plane.\footnote{With an abuse of notation, we use $z$ and $\phi$ in both one-point and two-point correlators. see Fig~\ref{fig:ecdiagram}.  Should be clear from context which case we refer to.} The two other coordinates will be integrated over.

\begin{figure}
	\centering
	\includegraphics[width=1\linewidth]{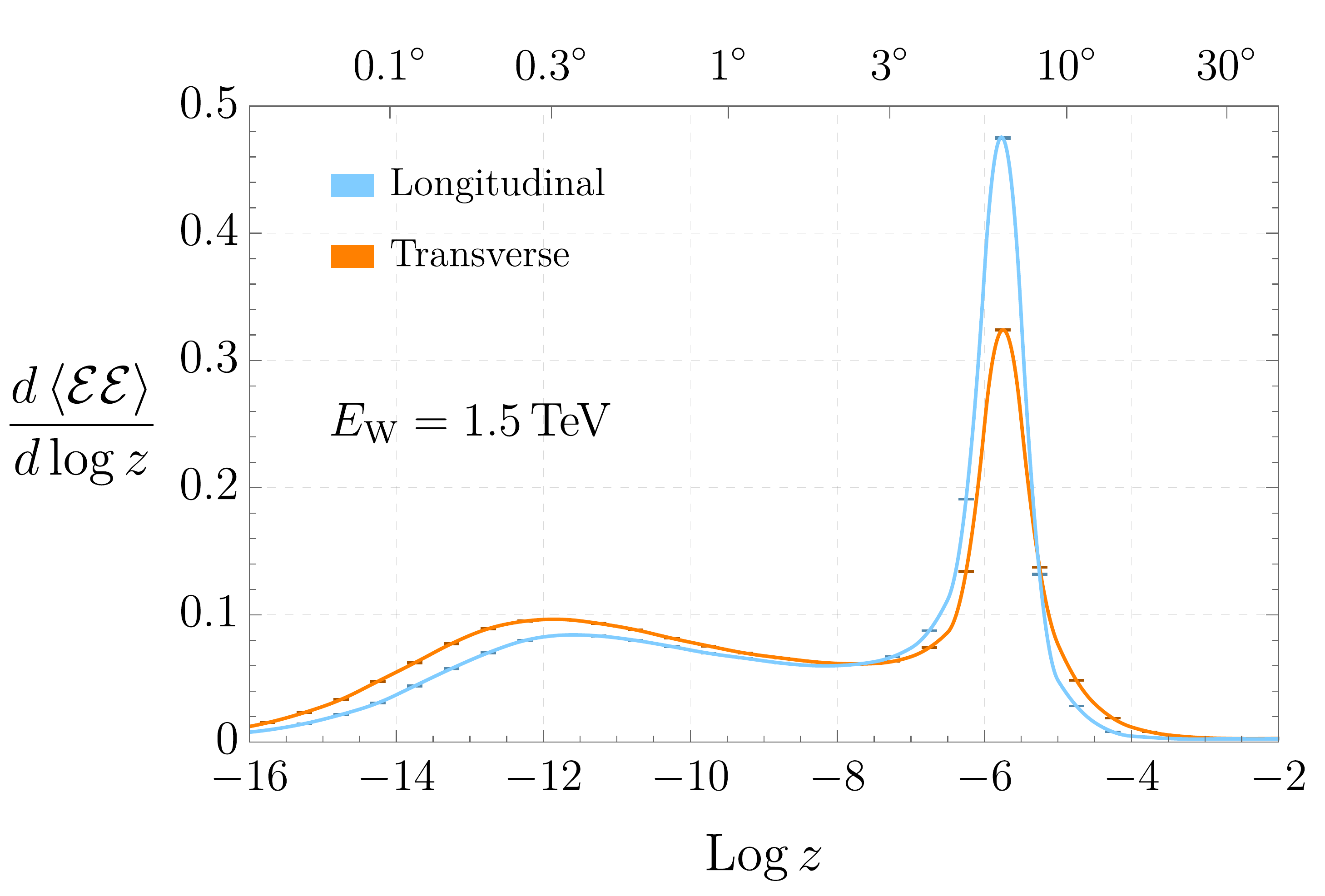}
	\caption{Two-point energy correlator of a $W$-boson jet. Both transverse and longitudinal distributions show a peak at the angular scale $z_\star=m_V^2/E^2$. For lower $z$, one explores the scaling regime of the individual quark sub-jets. At very small angular scales $z\sim \GeV^2/E^2$, hadronization enters and cuts off the correlator.}
	\label{fig:EEC}
\end{figure}

The amplitudes at leading order are given by Eq.~\eqref{eq:Ampsenfractions}, now with the energy fraction $x$ related to the angular separation $z$ between the quarks by 
$x(z)\,=\, \frac{1}{2}\left(1\pm \sqrt{1-z_\star/z}\right)$,
where the sign is determined by whether
$\cos\theta_*$ is positive or negative. 
At this order in the perturbative expansion, there is a minimal separation between quarks given by $z_\star$, equal for all vector polarizations. 
The product of the quark energies is given by $E_qE_{\bar{q}}/E^2=x(1-x)=\frac{z_\star}{4z}$, so the spectrum peaks at $z\sim z_\star$ for all entries of the density matrix. This is shown in Fig.~\ref{fig:EEC} for the diagonal ones. Since the longitudinal entry also behaves as $\propto x(1-x)$, the longitudinal peak is more pronounced than the one of the transverse distribution.

Two effects resolve the peak and give a contribution for lower $z$. First, the finite decay width of the vector gives a small width of order $\sim \Gamma_V/m_V$ to the peak in the energy correlator. However, this effect is subleading with respect the one given by the QCD radiation from the quarks. 
Indeed, the width of the peak corresponds to the angular scale where the correlator enters in the scaling regime. Such scaling regime 
has received recent attention \cite{Komiske:2022enw,Lee:2022ige} since it can be described as an operator product expansion of null-ray operators \cite{Hofman:2008ar,Kologlu:2019mfz}. For our purposes, it is sufficient to say that this regime is purely controlled by QCD, the $z$ dependence does not depend on the quark helicity, and apparently it does not offer any information that can be used to disentangle the different contributions of the diagonal entries of the decay density matrix.
Below the $\sim(\text{few}\GeV)^2/E^2$ scale, one can observe deviations due to hadronization \cite{Dokshitzer:1999sh,Belitsky:2001ij,deFlorian:2004mp}.

\medskip
\textit{Off-diagonal entries: Interference}.---
Both one- and two-point correlators of the decay density matrix depend on the azimuthal angles $\phi_{\langle \cale\rangle}$ and $\phi_{\langle \cale\cale\rangle}$, see Fig~\ref{fig:ecdiagram}, which we will denote generically $\phi$. The dependence is fixed to be $d\rho^V_{\hp h} \propto e^{i(h-\hp)\phi}$. If we use observables that are inclusive in $\phi$, like the invariant mass of the process or the $p_T$ of the vectors, only the diagonal terms with $h=\hp$ contribute while the off-diagonal ones with $h\neq \hp$ integrate to zero, giving the usual factorized picture of production rate times decay rate. 
Having control on the azimuthal angle $\phi$ allows to be sensitive to the interference between the production of different vector polarizations.

While this is inherently interesting, from a BSM perspective the attractiveness of being sensitive to interference terms is twofold. First, in the case of diboson production, while SM amplitudes are dominated by opposite-helicity transverse vectors, dimension six operators affect the production of either same-helicity vectors of longitudinally-polarized vectors. Therefore, interference terms between the dominant SM amplitudes and the BSM contribution have non-trivial $\phi$ dependence. Second, inclusive observables with quadratic sensitivity to EFT coefficients tend to be dominated by the high-energy tail of the distribution, which might induce problems with the EFT interpretation. Interference is linearly sensitive to higher dimension operators, allowing to interpret the constraints in a broader class of theories \cite{Farina:2016rws}. 

There are two types of interference depending on $\Delta h\equiv h-\hp$. Interference between the different transverse polarizations has $|\Delta h|=2$, while longitudinal-transverse interference has $|\Delta h|=1$. As mentioned previously, ignorance on whether the calorimeter is placed on the helicity-plus or helicity-minus quark amounts to a $x\to 1-x$ and $\phi\to \pi+\phi$ redundancy.
The effect of this redundancy on the $|\Delta h|=1$ interference will be discussed below. 
The effect on $|\Delta h|=2$ terms is simpler, since the phase $e^{i\Delta h\phi}$ is left invariant.

\begin{figure}
	\centering
	\includegraphics[width=1\linewidth]{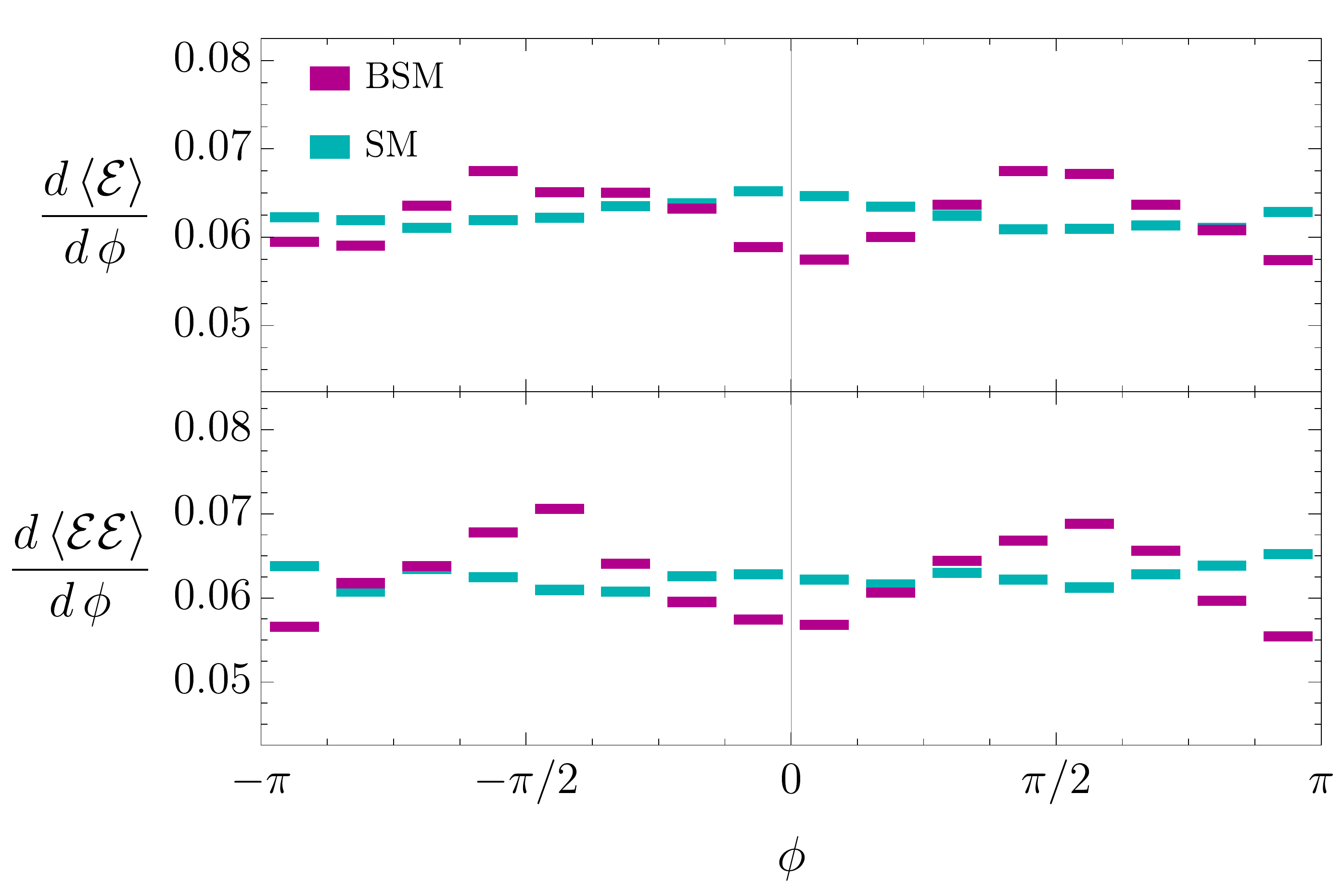}
	\caption{Azimuthal dependence of the one- and two-point energy correlators of a 1$\TeV$ $W$-boson.
	A contribution with $\lambda_\gamma=0.01$ shows a $\cos2\phi$ behaviour from the interference.}
	\label{fig:eceecphi}
\end{figure}

As an example, we study diboson production as a probe of such interference. At high energy, the SM produces mainly opposite-helicity vectors. Therefore $|\Delta h|=2$ interference of one vector should be studied along a $|\Delta h|=2$ interference of the other vector, making clear that interference is a phenomenon affecting the whole amplitude, and each process should be studied individually. However, in presence of the operator $\mathcal{L}\supset c_{WWW}\frac{v^2}{\Lambda^2}\frac{g^3}{m_W^2}\epsilon^{ABC}W^A_{\mu\nu}W^B_{\nu\rho}W^C_{\rho\mu}$, the triple gauge coupling discussed above is generated with $\lambda_\gamma=-6g^2c_{WWW}\frac{v^2}{\Lambda^2}$, giving an amplitude that produces same-sign-helicity $W$-bosons and grows at high energy. Therefore, a $|\Delta h|=2$ interference between the SM amplitude and the one driven by the EFT operator can be observed with a single vector while being inclusive on the kinematics of the other vector. This is shown in Fig.~\ref{fig:eceecphi}. We simulated the semileptonic decay of $e^+e^-\to W^+ W^-$ at $\sqrt{s}=2\TeV$ in the SM (cyan) and in presence of $\lambda_\gamma=0.01$ (magenta). The EFT operator induces a $\cos2\phi$ interfering pattern in both one- and two-point energy correlator of the $W$-boson jet. For the two point correlator, we impose the requirement $z_\star/2<z<2z_\star$ in order to avoid contributions from calorimeters separated a distance $z\ll z_\star$. The study of azimuthal dependence of the energy correlators at such small separations only reveals interference between different quark helicities, which is suppressed by the mass and of limited interest for the density matrix of a vector decay.\footnote{
Interference inside QCD jets has received some attention in the last years, see	
\cite{Chen:2020adz,Yu:2021zmw,Karlberg:2021kwr}.
}
We impose no constraint on $z$ in the one-point correlator.

\begin{figure}
	\centering
	\includegraphics[width=1\linewidth]{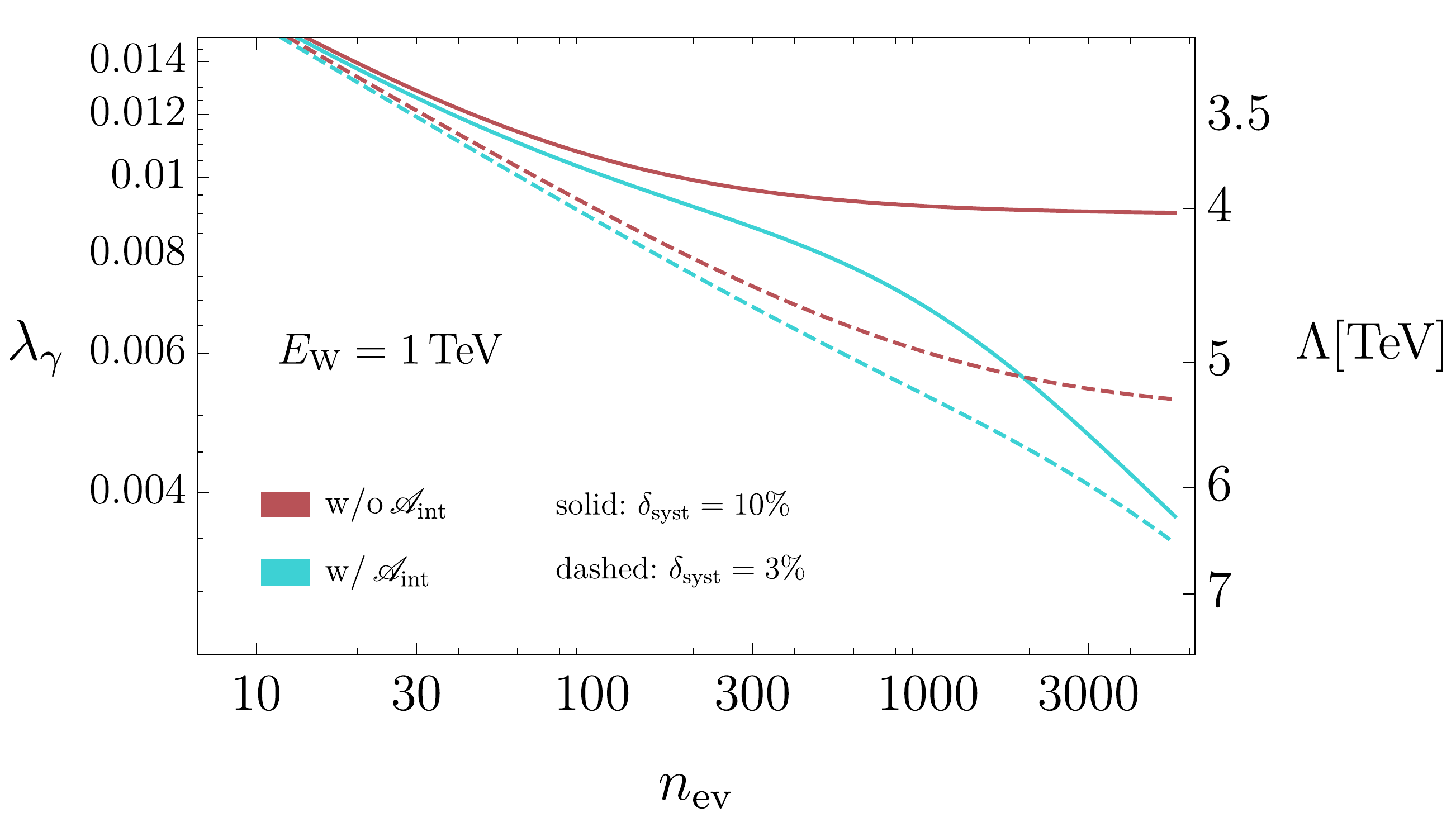}
	\caption{Impact of the asymmetry $\mathscr{A}_\text{int}$ on the projected 95\%CL sensitivity to $\lambda_\gamma$.}
	\label{fig:boundsphi}
\end{figure}

\begin{figure*}
	\centering
	\includegraphics[height=5.1cm]{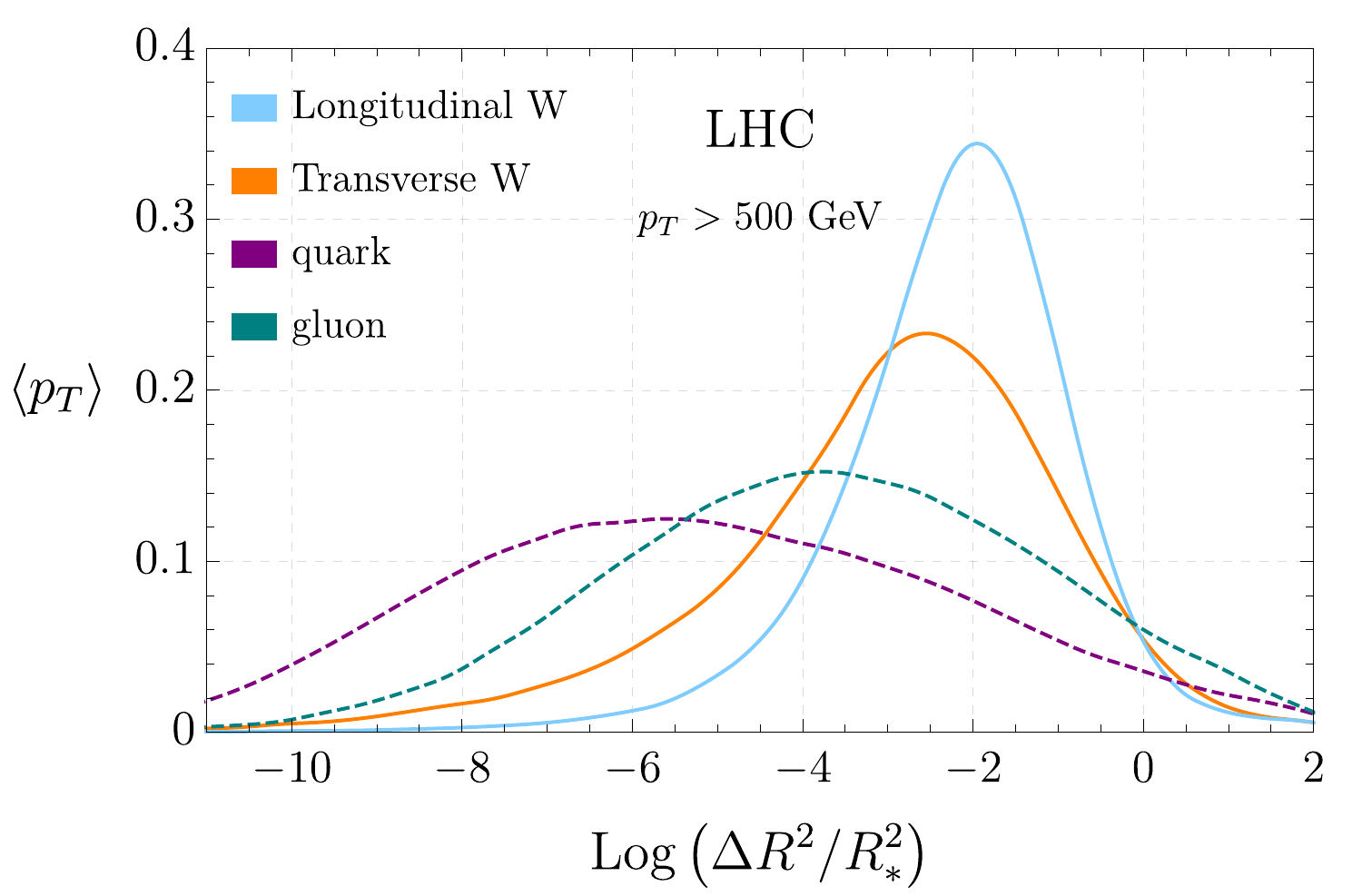}\hspace{0.5cm}
	\includegraphics[height=5.1cm]{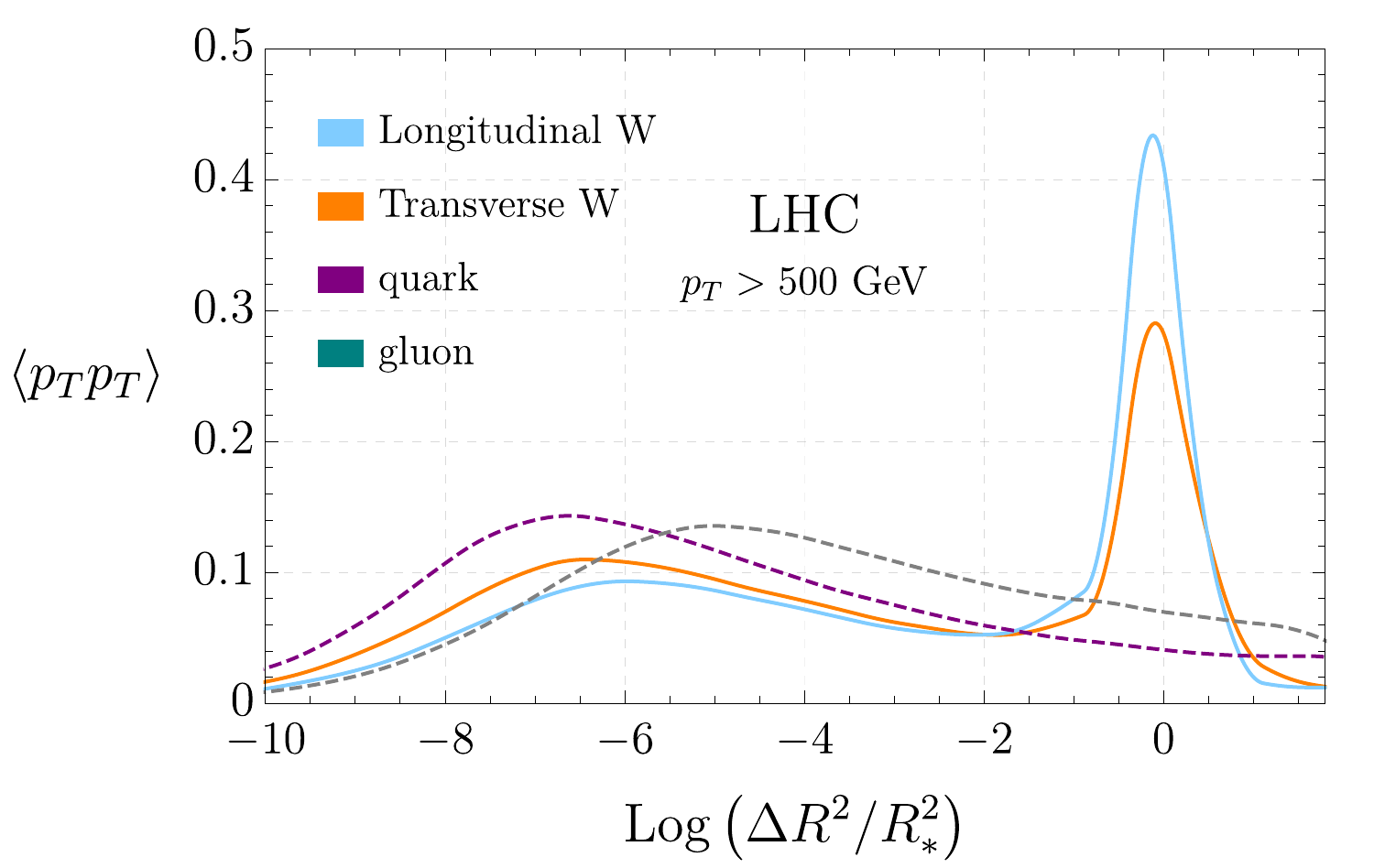}
	\caption{One-point (left) and two-point (right) $p_T$-correlators as a function of the boost invariant distance $\Delta R^2/R_\star^2$ at the LHC, for the transverse and longitudinal $W$ polarizations and a quark and gluon jet.}
	\label{fig:LHCcorrelator}
\end{figure*}
In order to have an estimate on how well the interference can be measured given a number of events, we design a simple observable, considering the accumulated energy in different quadrants of $\phi$. From a distribution $f(\phi)$ for the one- or two-point energy correlator, we define  $\mathscr{N}_+=f(\pi/4\leq |\phi|<3\pi/4)$ and $\mathscr{N}_-=f(0\leq |\phi|<\pi/4)+f(3\pi/4\leq |\phi|<\pi)$. Then the asymmetry $\mathscr{A}_\text{int}=(\mathscr{N}_+-\mathscr{N}_-)/(\mathscr{N}_++\mathscr{N}_-)$ is sensitive to the interference between different $W$-boson helicity production amplitudes.\footnote{Notice that $\mathscr{A}_\text{int}$, contrary to $\mathscr{A}_\text{LT}$, is in one-to-one correspondence to the one that can be defined at parton level.} 
For illustrative purposes we show in Fig.~\ref{fig:boundsphi} the bounds on $\lambda_\gamma$ as a function of the number of events at $\sqrt{s}=2\TeV$ with $p_T(W)>300\GeV$. In red we show the bounds from an inclusive measurement of the cross section assuming 10\% (solid) and 3\% (dashed) of systematic uncertainty. In cyan, we add the asymmetry $\mathscr{A}_\text{int}$ assuming negligible systematic uncertainty. For low number of events, the inclusive measurement saturates the sensitivity and the $\phi$ distribution gives no further discrimination, while for large number of events where precision measurements can be made, there is a clear advantage of using the asymmetry.  Moreover, the region where the asymmetry dominates scales faster with the number of events than the region dominated by the cross section measurement, since the interference term is linearly sensitive to $\lambda_\gamma$.
We've checked that if we only consider the terms up to $1/\Lambda^2$ in the cross section, adding the interference improves the bound by more than an order of magnitude, in accordance to \cite{Panico:2017frx,CMS:2021cxr}.
A detailed study of the full kinematics of the process allows to further improve the constraints, as done at parton level in \cite{Henning:2019vjr}.

We can now turn to the rest of Fig.~\ref{fig:2dconstraints}. In the right plot, we consider the same process and bin studied before in the left plot, but instead with $\lambda_\gamma=10^{-3}$ and $\delta g_{1z}=0$. Now it is the asymmetry $\mathscr{A}_\text{int}$ the one confirming that the excess comes indeed from an anomalous production of transverse modes. Moreover, due to the linear sensitivity, it can discriminate between both signs of $\lambda_\gamma$, contrary to the cross section measurement.

We stress that the conclusions above can be imported to LHC analyses: bins with large data samples will benefit the most from including observables sensitive to interference effects, since the linear sensitivity scales better with more statistics and allows to extend the regime of validity of the EFT.

We finish by commenting on the more challenging and, arguably, interesting $|\Delta h|=1$ interferences. As mentioned, the $x\to 1-x$ and $\phi\to \pi+\phi$ redundancy erases the interference pattern in $\phi$ if we proceed like for the $|\Delta h|=2$ interference. However, the interference can be observed by considering the dependence of both $\phi$ and $z$, since the $x\to 1-x$ changes the behaviour in $z$. This is equivalent as the necessity to observe two angles in the leptonic case as studied in \cite{Panico:2017frx,Chen:2020mev}. Therefore, the very interesting case of longitudinal-transverse interference is necessarily at least a two-dimensional problem. In practice, it is even more complex. In the case of diboson production, the SM production is dominated by opposite-sign transverse vectors while the typical BSM effects reside in the longitudinal-longitudinal production. Therefore, interference requires to observe a $|\Delta h|=1$ interference on both vector bosons, requiring a global study of all kinematics of the process. Such multidimensional problem is better suited for modern machine learning approaches \cite{Baldi:2016fzo,Brehmer:2018kdj,Brehmer:2018eca,Brehmer:2019xox,Wunsch:2020iuh,Chen:2020mev}.

\medskip
\textit{Towards the LHC}.---
So far we have studied the energy correlators for a monoenergetic source of electroweak bosons. In a hadronic environment, as the LHC, one has a spectrum of energies. However, the previous theoretical predictions can be immediately imported due to two effects. 
First, it is possible to show that the energy ratio $E_i/E_J$ between a hadron and jet is related to the ratio of their transverse momenta,
$E_i/E_J=p_{T,i}/p_{T,J}+\mathcal{O}(z_\star^{1/2})$. Moreover, the angular distance $z$ is related to the boost invariant distance $\Delta R^2=(\Delta\eta)^2+(\Delta\phi)^2$ by
$z/z_\star=\Delta R^2/R_\star^2+\mathcal{O}(z_\star^{1/2})$, where we have defined $R_\star=4m_V^2/p_{T,J}^2$.
Second, at leading order in $z_\star$, the LO predictions only depend on the quantity $z/z_\star$. This way, by measuring $z_\star$ for each event, the obtained distribution from an ensamble of events in terms of $z/z_\star$ will look like a distribution from a monoenergetic source.
Therefore, the boost invariant $p_T$-\textit{correlator}, where we weight by the $p_T$ of the hadrons as a function of the distance $(\Delta R)^2/R_{\star}^2$, has the same distribution as the energy correlator up to $\mathcal{O}(z_\star^{1/2})$ corrections.

In Fig.~\ref{fig:LHCcorrelator} we show the one- and two-point $p_T$-correlator as a function of $(\Delta R)^2/R_{\star}^2$ for a longitudinal and transverse $W$-boson, and for reference we also for a quark and gluon jet. The jets are reconstructed using the anti-$k_t$ algorithm \cite{Cacciari:2008gp} with $R=1$. The distributions for the vector coincide with the predictions for the EC of a monoenergetic $W$ beam, at least up to $\mathcal{O}(z_\star^{1/2})$ corrections. The quark and gluon distributions for the one-point peak at lower $(\Delta R)^2/R_{\star}^2$ as expected.

Moreover, notice that the one-point correlator requires the definition of a jet axis, which can be sensitive to soft recoil. It would be interesting to study the one-point correlator with a stable definition for the jet axis \cite{Larkoski:2014uqa,Marzani:2019hun}.

The two-point shows a completely different behaviour between EW and QCD jets. While the vector distributions peak at $z_\star$, the quark and gluon jets show the scaling behaviour.  In fact, the tail at low $(\Delta R)^2/R_{\star}^2$ of the vector distributions coincides with the quark-jet distribution, as expected. 

We conclude that the previous monoenergetic results can be imported for the LHC distributions for the diagonal entries of the density matrix.
The azimuthal distribution of the off-diagonal entries receives, however, modifications due to boosts along the beam axis. While we expect mild effects, 
the assessment of the sensitivity requires further study. 

\begin{figure}
	\centering
	\includegraphics[width=1\linewidth]{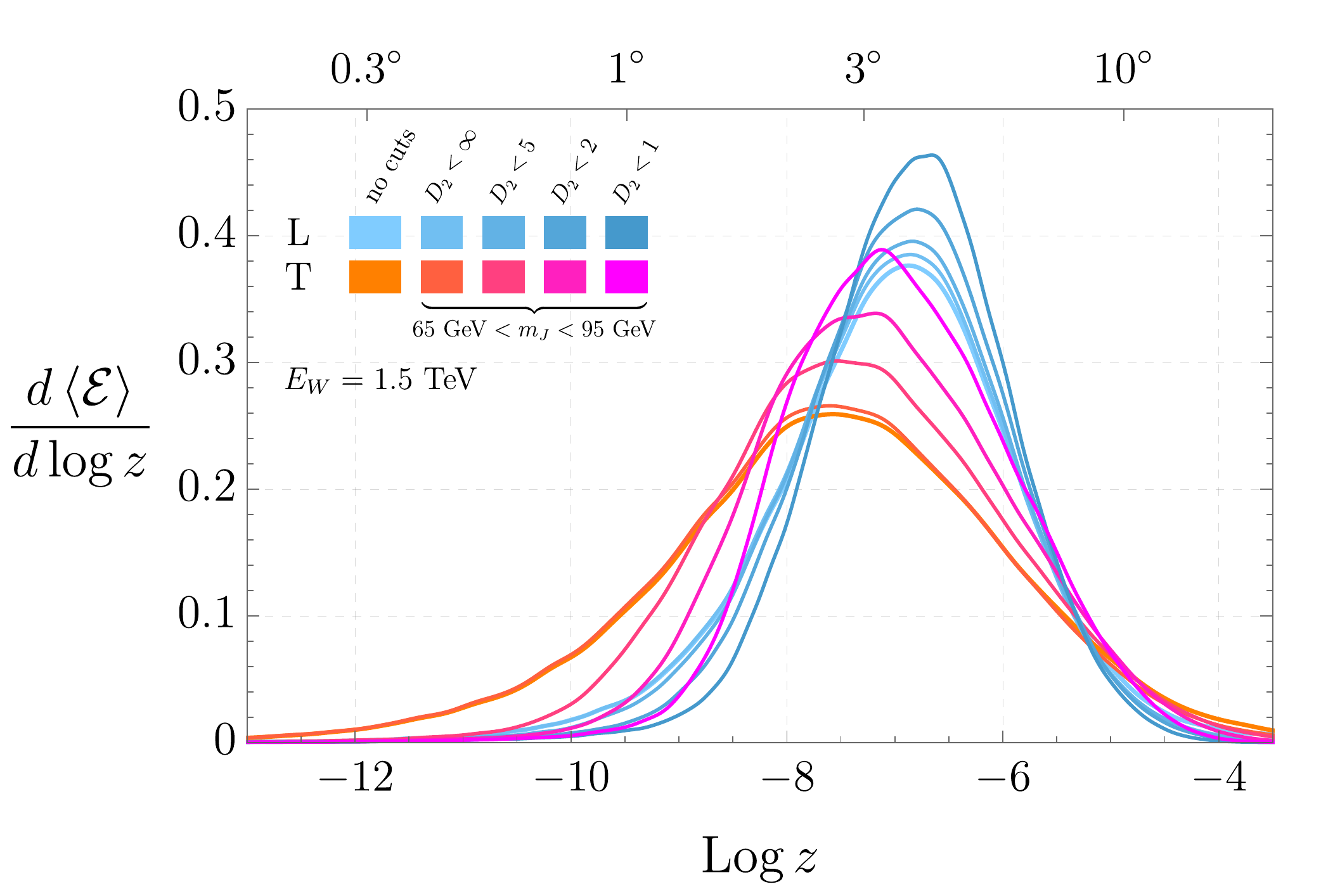}
	\caption{Impact on the one-point correlator of the selection cuts typically used to discriminate EW from QCD jets.}
	\label{fig:ECselectioncuts}
\end{figure}
\medskip
An important question is the impact of the QCD background and how the spectrum of the energy correlators is deformed after applying the selection criteria typically used to discriminate EW and QCD jets \cite{Thaler:2010tr,Larkoski:2014gra,Larkoski:2015kga,Larkoski:2017iuy,ATLAS:2019nat}. In particular, the invariant mass of the jet $m_J$ and the $D_2$ observable \cite{Larkoski:2015kga} are two of the main drivers that discern between EW and QCD jets.
In order to assess the impact of such cuts in a clean environment, we consider a monoenergetic beam of $W$ bosons of $E_W=1.5\TeV$ in a lepton collider. 
The effect of the cuts on the one-point correlator is shown in Fig.~\ref{fig:ECselectioncuts}, where we compare the spectrum without cuts, after the requirement $65\GeV<m_J<95\GeV$, and after further requiring $D_2<5$, 2 and 1. All curves are normalized to 1.
We observe that the cut on $m_J$ has only mild effects on the spectrum.
However, a selection on $D_2$ can have a large impact. 
The $D_2$ observable is defined as
\be
D_2 = \frac{\frac{1}{E_J^3}\sum_{i<j<k}E_iE_jE_k\,z_{ij}z_{jk}z_{ki}}{\left(\frac{1}{E_J^2}\sum_{i<j}E_iE_j\,z_{ij}\right)^3},
\ee
where the sums run over the jet constituents. Low values of $D_2$ correspond to jets that look like two-prong while larger values correspond to one-prong jets,  in this way cut selections can be designed to reject gluonic jets.
In Fig.~\ref{fig:ECselectioncuts} we see, however, that such selections have an inherent bias towards rejecting the transverse vectors that are kinematically distinct with respect the longitudinals. 
This is because, as explained in the discussion of the one-point correlator, low $z$ is in one-to-one with having the energy deposition dominated in a single quark, which translates to larger $D_2$ values. Such kinematical configuration is more frequent in the transverse polarizations. The $D_2$ cut imposes a requirement on the kinematical configuration within the jet and homogenizes the transverse and longitudinal distributions.

We cannot end the discussion without remarking that $D_2$ is defined as a ratio of fully integrated energy correlators, and therefore it is suggestive to think that the exploration of fully-differential observables might allow a discrimination of EW and QCD jets that retains the different character of longitudinal and transverse vectors.

\medskip
\textit{\textbf{Conclusions and outlook}}

In this \textit{Letter} we have explored the possibility of using the information encoded in the energy correlators of a hadronically decaying electroweak vector boson in order to probe its full decay density matrix. We observe that the dependence on the angular separation $z$ of the one-point energy correlator can indeed discriminate between longitudinal and transverse modes. The azimuthal dependence $\phi$ alone of both one- and two-point energy correlators shows the interference pattern of the $|\Delta h|=2$ interference terms. The $|\Delta h|=1$ interference terms require the study of the distribution in both $\phi$ and $z$. The distributions are independent of the $W$ boost when expressed as a function of $z/z_\star$, which allows to predict the distributions observed at hadron colliders, like the LHC, in terms of boost invariant quantities.

Kinematics of the decay allows to dissect the dynamics that produce an electroweak boson, enhancing the sensitivity to EFT operators and potentially to resonance searches.
We presented how the energy correlators allow to identify the microscopic origin of an excess of events, see Fig.~\ref{fig:2dconstraints}, 
and how the sensitivity to EFT operators is enhanced by unveiling the interference pattern, see Fig.~\ref{fig:boundsphi}.

An important future avenue to pursue is the detailed analysis of the correlators in the hadronic environment of the LHC. 
The impact of QCD jets and selection criteria on the distributions is a crucial element to explore further. We observe that the usual $D_2$ selection cut has a bias towards eliminating vectors with transverse-like kinematics, forcing the exploration of other criteria that may retain the kinematical difference between the longitudinal and transverse distributions.
This, together with the assessment of the different theoretical and experimental uncertainties is postponed for future work. 

While we studied the one-dimensional distributions of the energy correlator, extra information can be gained from the multi-dimensional distributions.
For instance, the full 4-dimensional two-point and 6-dimensional three-point correlator would constitute a generalization of the $D_2$ variable used to discriminate QCD and EW jets, which might be interesting to investigate.

From a BSM perspective, a better control of hadronic decays and the kinematical characterization of the decaying EW vectors might open the door to study rare multiboson processes that carry important information about the microscopic dynamics in the EW sector \cite{Henning:2018kys}.

Energy correlators are at the crossroads of conformal field theory, QCD, experimental physics and data analysis. 
As shown, they can also be used to enhance sensitivity to BSM physics.
This makes them an extremely appealing and fruitful arena that aims to push precision physics at the LHC to new levels.

\subsection*{Acknowledgments}

We thank Alfredo Glioti, Ian Moult, Francesco Riva, Steven Schramm and Andrea Wulzer for valuable discussions, suggestions and comments.
The work of L.R. and M.R. is supported by The Swiss National Science Foundation under contract 200021-178999 and 200020-188671. 

\appendix

\bibliography{bibs} 

\end{document}